\documentclass[10pt,twocolumn,aps,pra,superscriptaddress,showpacs,tightenlines,pdflatex,longbibliography]{revtex4-2}

\usepackage{amsthm}
\usepackage{gensymb}
\usepackage{amsmath}            %serve per le subequazioni
\usepackage{amssymb}            %serve per il simbolo "marchio registrato", \circledR
\usepackage{mathtools}
\usepackage{graphicx}           %serve per le figure eps, ps etcyy
\graphicspath{ {./images/} }
\usepackage{xcolor}

\usepackage{etoolbox}
\usepackage{breqn}

\usepackage{xcolor}
\usepackage{bm}
\makeatletter
\let\cat@comma@active\@empty

\usepackage[colorlinks,citecolor=blue,urlcolor=blue,bookmarks=false,hypertexnames=true]{hyperref} 
\newcommand{\ket}[1]{| #1 \rangle}
\newcommand{\bra}[1]{\langle #1 |}

\newcommand{\red}[1]{{\color{black} #1}}

\newcommand{\comment}[1]{}

\begin{document}
\title{Hidden nonmacrorealism: reviving the Leggett-Garg inequality with stochastic operations}

\author{Huan-Yu Ku}
\email{These authors contributed equally.}
\affiliation{Department of Physics and Center for Quantum Frontiers of Research \&
Technology (QFort), National Cheng Kung University, Tainan 701, Taiwan}
\affiliation{Theoretical Quantum Physics Laboratory, RIKEN Cluster for Pioneering Research, Wako-shi, Saitama 351-0198, Japan}
\author{Hao-Cheng Weng}
\email{These authors contributed equally.}
\affiliation{Department of Physics, National Tsing Hua University, Hsinchu 30013, Taiwan\\
and Center for Quantum Technology, Hsinchu 30013, Taiwan}
\author{Yen-An Shih}
\affiliation{Department of Physics, National Tsing Hua University, Hsinchu 30013, Taiwan\\
and Center for Quantum Technology, Hsinchu 30013, Taiwan}
\author{Po-Chen Kuo}
\affiliation{Department of Physics and Center for Quantum Frontiers of Research \&
Technology (QFort), National Cheng Kung University, Tainan 701, Taiwan}
\affiliation{Theoretical Quantum Physics Laboratory, RIKEN Cluster for Pioneering Research, Wako-shi, Saitama 351-0198, Japan}
\author{Neill Lambert}
\affiliation{Theoretical Quantum Physics Laboratory, RIKEN Cluster for Pioneering Research, Wako-shi, Saitama 351-0198, Japan}
\author{Franco Nori}
\affiliation{Theoretical Quantum Physics Laboratory, RIKEN Cluster for Pioneering Research, Wako-shi, Saitama 351-0198, Japan}
\affiliation{RIKEN Center for Quantum Computing, Wako, Saitama 351-0198, Japan}
\affiliation{Department of Physics, The University of Michigan, Ann Arbor, 48109-1040 Michigan, USA}
\author{Chih-Sung Chuu}
\email{cschuu@phys.nthu.edu.tw}
\affiliation{Department of Physics, National Tsing Hua University, Hsinchu 30013, Taiwan\\
and Center for Quantum Technology, Hsinchu 30013, Taiwan}
\author{Yueh-Nan Chen}
\email{yuehnan@mail.ncku.edu.tw}
\affiliation{Department of Physics and Center for Quantum Frontiers of Research \&
Technology (QFort), National Cheng Kung University, Tainan 701, Taiwan}
\affiliation{Theoretical Quantum Physics Laboratory, RIKEN Cluster for Pioneering Research, Wako-shi, Saitama 351-0198, Japan}

\begin{abstract}
  The Leggett-Garg inequality (LGI) distinguishes nonmacrorealistic channels from macrorealistic ones by constraining the experimental outcomes of the underlying system. In this work, we propose a class of channels which, initially, cannot violate the LGI (in the form of the temporal Bell inequality) but can violate it after the application of stochastic pre- and post- operations (SPPOs). As a proof-of-principle experiment, we demonstrate the stochastic pre- and post- operations in an amplitude-damping channel with photonic qubits. We denote the above phenomenon as hidden nonmacrorealistic channels. We also discuss the relationship between this hidden nonmacrorealistic channels (in terms of the temporal Clauser-Horne-Shimony-Holt (CHSH) inequality) and the strongly nonlocality-breaking channel, which breaks the hidden spatial CHSH nonlocality for arbitrary states. In general, if the channel satisfies hidden nonmacrorealism, it is not a strongly CHSH nonlocality-breaking channel.  %Our results advance research on temporal quantum correlation with the quantum communication. \red{HY: Maybe above is unnecessary. \red{nl: yeah maybe better in the intro to make these statements}}
\end{abstract}

\maketitle

\section{Introduction}\label{sec:intro}
The location of the boundary~\cite{Schrodinger35} between classical macroscopic systems, which have definite or pre-existing physical properties, and microscopic quantum objects, obeying the uncertainty principle, is still under debate. %For providing a counterintuitive example for the macroscopic system, the famous gedanken experiment is used to explore whether the Schr\"odinger's cat is dead or alive~\cite{Schrodinger35}. 
%The famous gedanken experiment of Schr\"odinger's cat is often used for the discussion of this debate~\cite{Schrodinger35}. 
In the seminal work~\cite{Leggett85}, Leggett and Garg provided a quantitative way to broach this problem with an inequality which distinguishes genuine quantum superpositions from macroscopic systems with definite (macrorealistic) observables. %e.g., mass, length-scale, for a number of the particles constituting the system is considering~\cite{Leggett85}. 
Violating the Leggett-Garg inequality (LGI) implies that the dynamics of the system is nonmacrorealistic, which means the experimental outcomes violate either macrorealism \textit{per se} or noninvasive measurability (or both)~\cite{Emary2014Review}. Without loss of generality, the dynamics of the system, which is used to construct the two-time correlation function for computing the LGI, can always be described by quantum channels. Therefore, when there is no ambiguity, throughout this work, we will use the terminology of quantum channels instead of the dynamical process of the system. 

Based on the definition of macrorealism, variations and refinements of the original LGI can be derived, including the  quantum witness (or no-signaling-in-time condition)~\cite{Li2012, Kofler13,Clemente2016,Halliwell2016,Halliwell2017}, temporal Bell inequalities~\cite{Brukner2004,Fritz2010}, transport-based inequalities~\cite{Lambert10,Lambert102,Emary12}, higher dimensional LGIs~\cite{Budroni14,Moreira15,Lambert16}, multi-time correlations~\cite{Ringbauer2018,Uola2019,Berk2021resourcetheoriesof}, and \red{the partial exclusion of false violations arising from clumsy (classi-
cally invasive) measurement~\cite{Wilde2011,Knee16,Emary2017,Moreira2019,Ku2019}.} 
In addition, the range of such tests can be presented as a hierarchy of temporal correlations~\cite{YuehNan2014,Che-Ming2015,Ku2018-2,Uola2018}. Furthermore, many significant experiments have been implemented in different systems, such as photons~\cite{Goggin2011,Dressel2011,Zhou2015,Wang2020}, superconducting qubits~\cite{PalaciosLaloy2010,Knee2012,White2016,Knee16,Huffman17,Ku2019}, and atoms~\cite{Budroni15,Robens15}. Recently, temporal quantum correlations (including LGI) have found applications for quantum information and communication problems, like witnessing the dimension of a system~\cite{Schild2015,Hoffmann_2018,Spee_2020}, the certification of quantum memory~\cite{Rosset2018,Budroni_2019,Uola2020,Langenfeld2020,Vieira2021,Yuan2021}, quantum clock~\cite{Budroni2021}, enhancing the quantumness of a system with non-Markovianity~\cite{Wu2020,Milz2020}, self-testing the quantum measurement~\cite{Maity2021}, and verifying the quantumness of a channel~\cite{Pusey2015,Wang2019,Lin2020,Shih-Hsuan2020,Ku2021}.

In this work, we propose a class of quantum channels that alone cannot violate the temporal Bell inequality. However, after applying stochastic pre- and post-operations (SPPOs), this class of channels can violate the temporal Bell inequality. We denote such channels as ``hidden nonmacrorealistic channels'', akin to ``hidden nonlocality'' in the context of the spatial Bell inequality~\cite{Bell64,CHSH1969,Brunner2014RMP}, in which a local state can be activated to become nonlocal by stochastic local operations and classical communication~\cite{Popescu1995,Gisin1996,Masanes2008,Hirsch2013,Wang2020-2}. In comparing these two similar scenarios, we find an additional implication of our result for  {\em strongly} Clauser-Horne-Shimony-Holt  (CHSH) nonlocality-breaking channels, which break the hidden CHSH nonlocality for arbitrary states~\cite{Pal2015}. We show that when the channel satisfies hidden nonmacrorealism, it is not a strongly CHSH nonlocality-breaking channel. Therefore, the LGI can be applied in the certification of the quantumness in a quantum network.

This work is organized as follows. In Sec.~\ref{sec:PRELIMINARY NOTIONS}, we introduce the formalism and basic concepts related to macrorealism and strongly CHSH nonlocality-breaking channel. In Sec.~\ref{sec:Hidden nonmacrorealism}, we introduce the concept of hidden nonmacrorealism under the temporal CHSH scenario and its relation to the strongly CHSH nonlocality-breaking channel. In Sec.~\ref{Sec_Experimental_demonstration}, we experimentally demonstrate the hidden nonmacrorealism in an amplitude-damping channel. Finally, in Sec.~\ref{sec:conclusion}, we present the conclusion and outlook of our work.

\section{PRELIMINARY NOTIONS}\label{sec:PRELIMINARY NOTIONS}
\subsection{Temporal Clauser-Horne-Shimony-Holt inequality}

In this section, we briefly recall the Leggett-Garg inequality (LGI) in the form of the temporal Bell inequality~\cite{Emary2014Review, Fritz2010, Brukner2004}. We consider a system undergoing sequential measurements with classical inputs $x (y)$ and outcomes $a (b)$ acting at time $t_0$ ($t_1$). A schematic diagram of this process is shown in Fig.~\ref{Schematic}. If the dynamics of the system obeys macrorealism {\it per se} (the system is always in a macroscopic state), and non-invasive measurability (measurements only reveal the physical property of the state without disturbance~\cite{Emary2014Review,Leggett85}), the observed probability distribution for the outcomes $\{p(a,b|x,y)\}$ must be macrorealistic. That is,
\begin{equation}\label{Eq_LHV}
p(a,b|x,y)\stackrel{\text{MS}}=\sum_{\lambda}p(\lambda)p(a|x,\lambda)p(b|y,\lambda),
\end{equation}
%We can easily see that the above formula formulating the famous ``The god does not throw dice" because.
where $\lambda$ is the hidden parameter which determines causally all physical properties including the probability distributions $p(a|x,\lambda)$ and $p(b|y,\lambda)$.% Here, macrorealism {\it per se} and non-invasive measurability respectively assume that the system always be in a macroscopic state and measurements only reveal the physical property of the state without changing it.

In contrast, employing quantum theory, the observed probability distribution can be formulated by Born's rule, namely
\begin{equation}\label{Eq_pabxy}
p(a,b|x,y)\stackrel{\text{QM}}= \text{Tr}\left[M_{b|y}\mathcal{E}(M_{a|x}\rho_0 M_{a|x})\right],~\forall a,b,x,y.
\end{equation}
Here, $\rho_0$ is the initial state, $\mathcal{E}$ is a quantum channel describing the dynamics between time $t_0$ and $t_1$, and $\{M_{a|x}\}$ ($\{M_{b|y}\}$) is the von-Neumann measurement set. %Note that the subnormalized post-measurement states $\sqrt{M_{a|x}}\rho_0\sqrt{M_{a|x}}~\forall~a,x$ are described by the von Neumann-L\"uders instrument~\cite{Lders1950} by introducing the Kraus operator $K_{i|j}=\sqrt{M_{i|j}}$.
It has been shown that there exist statistics (generated by quantum theory) not admitting Eq.~\eqref{Eq_LHV}~\cite{Emary2014Review, Fritz2010, Brukner2004,Waldherr2011,Majidy2021}. Thus, we denote such a correlation in time as \textit{nonmacrorealism}. 

The assumption of non-invasive measurability can be recast as the so-called no-signalling-in-time condition~\cite{Kofler13,Clemente2016,Li2012,Halliwell2017,Halliwell2016,Halliwell2019}, 
\begin{equation}\label{Eq_no_signalling_in_time}
\sum_{a}p(a,b|x,y)=\sum_{a}p(a,b|x',y)~~\forall~x\neq x'.
\end{equation}
The above formula indicates that the choice of the measurement at time $t_0$ does not ``disturb" the observing probability distribution at time $t_1$.
%it is not possible to detect from the observed statisticswhether a measurement has been performed
%The physical interpretation of the above formula is that the marginals of the statistics at time $t_0$ are independent of the choice of the measurement setting $x$.
To satisfy Eq.~\eqref{Eq_no_signalling_in_time} for an arbitrary measurement set $\{M_{a|x}\}$, we can consider the initial state $\rho_0$ to be the maximally mixed state because the unnormalized post-measurement state for von-Neumann measurement can be described by $M_{a|x}\rho_0 M_{a|x}$.  %\red{Nl: please clarify} We refer to the generalized case, in which the no-signaling-in-time condition is considered, to be the non-disturbance problem in Ref.~\cite{Uola2019}. \red{Although throughout this work we only consider von-Neumann measurements in the temporal Bell inequality, the general non-disturbance condition can be trivially applied.}
%We note that the non-invasive measurability assumption can be partially addressed by the use of clumsy-measurement invasiveness tests with additional control experiments~\cite{Knee16,Wilde2011,Emary2017,Moreira2019,Ku2019}.

To demonstrate our main result, it is convenient to introduce the simplest temporal Bell inequality with the index sets $x,y\in\{1,2\}$ and $a,b\in\{\pm 1\}$. %\red{Inserting the causality, in which the information only propagates forward in time [e.g., Eq.~\eqref{Eq_no_signalling_in_time}], into the definition of the macrorealism,} The temporal Clauser-Horne-Shimony-Holt (CHSH) inequality can be obtained~\cite{Fritz2010} as
The temporal Clauser-Horne-Shimony-Holt (CHSH) inequality is~\cite{Fritz2010}
\begin{equation}\label{Eq_CHSH}
B_{\text{T-CHSH}}=C_{1,1}+C_{2,1}+C_{1,2}-C_{2,2}\leq 2,
\end{equation}
where $2$ is the macrorealistic bound, and $C_{x,y}=\sum_{a,b} a \cdot b~p(a,b|x,y)$ is the two-time correlation function. \red{Here, the indices ``a" and ``b" represent the measurement outcomes of Alice and Bob with measurement bases labeled by $x$ and $y$ at different times $t_0$ and $t_1$, respectively. If the temporal CHSH inequality is not violated, the correlation is regarded as CHSH macrorealistic.}
We note that there is no difference between the temporal CHSH and the standard CHSH inequality unless one a priori knows how to generate the probability distributions.

%$\{M_{a|x}\} (\{M_{b|y}\})$

%The physical interpretation of the above equation is the follow-ing: The probability distributionp(a, b|x, y) between timest1andt2does not depend on the history of the experiment.Therefore, there exist hidden parametersλ, which can be deter-ministic or stochastic \cite{Clemente2015}, defining all physical properties andforming the probability distributionsp(a|x, λ) andp(b|y, λ)

%an initial system $\rho_{0}$ is measured by sequential measurements $\{M_{a|x}\} (\{M_{b|y}\})$ with classical inputs $x (y)$ and outcomes $a (b)$ 

\subsection{Strongly Clauser-Horne-Shimony-Holt nonlocality-breaking channel}
\red{Before showing our main result, we briefly recall the concept of the (strongly) CHSH nonlocality-breaking channel~\cite{Pal2015,Zhang2020}. A bipartite state $\rho_{\text{AB}}$ is said to be CHSH nonlocal if it violates the CHSH inequality, which has the same form of Eq.~\eqref{Eq_CHSH}. Otherwise, we say the state is CHSH local. In addition, we say that a state is hidden CHSH nonlocal if a CHSH local state can become nonlocal after applying stochastic local operations with classical communication. Here, we use the local filtering operations, denoted by $\Lambda_{\rm A}$ and $\Lambda_{\rm B}$, to implement the stochastic local operations with classical communication~\cite{Gisin1996,Masanes2008,Hirsch2013}. A bipartite state after local filters can be described by
\begin{equation}\label{eq:state_after_local_filter}
    \rho'_{\rm AB}= (K_{\rm A}\otimes K_{\rm B})(\rho_{\rm AB})(K_{\rm A}^\dagger\otimes K_{\rm B}^\dagger)/N,
\end{equation}
where $K_{\rm A}$ ($K_{\rm B}$) is the Kraus representation of the local filtering operation, and $N=\text{Tr}\left[(K_{\rm A}\otimes K_{\rm B})(\rho_{\rm AB})(K_{\rm A}^\dagger\otimes K_{\rm B}^\dagger)\right]$ is the normalization constant. In general, a local filtering operation is a complete-positive (CP) trace-nonincreasing map with only one Kraus operator. Therefore, we can define another Kraus operator, $K_2=\sqrt{\openone-K_1^\dagger K_1}$, such that $K_1^\dagger K_1+K_2^\dagger K_2=\openone$ forms a quantum channel. In addition, since there are only two Kraus operators in the above quantum channel, the system through the local filter ($K_1$) can be regarded as a successful operation. On the other hand, the system after $K_2$ is denoted as a fail operation. In this case, $N$ represents the success probability.

Now, if a correlation, obtained after the channel $\mathcal{E}$ acting on the subsystem, always satisfies the CHSH inequality for arbitrary measurements and bipartite states, the channel is CHSH noncality-breaking. More specifically, a CHSH nonlocality-breaking channel generates the correlation
\begin{equation}\label{Eq_spatial_statistic}
p(a,b|x,y)=\text{Tr}\left[M_{a|x}\otimes M_{b|x} (\openone\otimes\mathcal{E})\rho_{\text{AB}}\right],
\end{equation}
which always satisfies the CHSH inequality with arbitrary measurement sets $\{M_{a|x}\}$ ($\{M_{b|y}\}$), and state $\rho_{\text{AB}}$. Note that the measurement $M_{a|x}(M_{b|y})$ is implemented on one of the spatially separated states.

Furthermore, if the evolved state $(\openone\otimes \mathcal{E})\rho_{\text{AB}}$ does not show hidden CHSH nonlocality for any input state and any choice of measurements, %the renormalized quantum state $(\Lambda_{\text{A}}\otimes \Lambda_{\text{B}})(\openone\otimes \mathcal{E})\rho_{\text{AB}} (\Lambda_{\text{A}}\otimes \Lambda_{\text{B}})^\dagger/N$ with $N=\text{Tr}[\Lambda_{\text{A}}^\dagger \Lambda_{\text{A}}\otimes \Lambda_{\text{B}}^\dagger\Lambda_{\text{B}}(\openone\otimes \mathcal{E})\rho_{\text{AB}} ]$ cannot violate the CHSH inequality. 
the channel can be further denoted as a {\em strongly CHSH nonlocality-breaking channel}. It has been shown~\cite{Pal2015} that the strongly CHSH nonlocality-breaking channel can be certified by testing whether the Choi state of the channel, defined as $\rho_{\text{CJ}}=(\openone\otimes \mathcal{E})\ket{\Phi_+}\bra{\Phi_+}$ with $\ket{\Phi_+}=1/\sqrt{d}\sum_i\ket{i}\otimes \ket{i}$, shows the hidden CHSH nonlocality. Note that the Choi state can be defined not only for the quantum channel but also for CP maps~\cite{Bavaresco2019semidevice,Abbott2020communication,Giarmatzi2021witnessingquantum}.}

\section{Hidden nonmacrorealism}\label{sec:Hidden nonmacrorealism}
\begin{figure}[tbp]
\includegraphics[width=1\columnwidth]{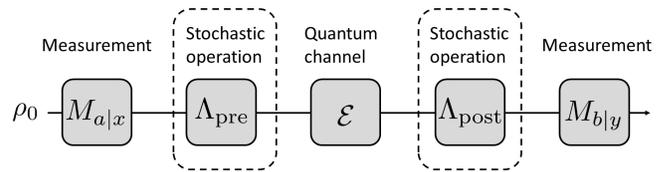}
\caption{Schematic diagram. Without the process in the dashed-line boxes, the schematic diagram represents the standard nonmacrorealism test. Hidden nonmacrorealism can be observed by introducing the stochastic operations in the dashed-boundary boxes.}
\label{Schematic}\centering 
\end{figure}

We now consider a more general protocol by which we can ``activate" nonmacrorealism from apparently macrorealistic channels by introducing stochastic pre- and post-operations (SPPOs). %\red{NL: reword the following: acting on the channels in the Heisenberg picture}. 
The schematic diagram of the process is shown in Fig.~\ref{Schematic}. More precisely, given a quantum channel %\red{NL: process? channel? propagator?}
$\mathcal{E}$, that obeys Eq.~\eqref{Eq_LHV} for arbitrary measurements $\{M_{a|x}\}$ and $\{M_{b|y}\}$, nonmacrorealism can be observed after applying the SPPOs described by the pre and post filters $\Lambda_{\text{pre}}$ and $\Lambda_{\text{post}}$. The probability distribution of the outcomes from the whole process including measurements, SPPOs, and channel can now be formulated as 
\begin{equation}\label{Eq_pabxy_instrument}
p(a,b|x,y)=\text{Tr}\left[M_{b|y}\left(\frac{\Lambda_{\text{post}}\circ\mathcal{E}\circ\Lambda_{\text{pre}}(M_{a|x})}{d\times N(a|x)}\right)\right],
\end{equation}
where the normalization constants are
\begin{equation}\label{Eq_normalzation_constant}
N(a|x)=\text{Tr}\left[\Lambda_{\text{post}}\circ\mathcal{E}\circ\Lambda_{\text{pre}}(M_{a|x})\right]/d.
\end{equation}
Here, the initial input with dimension $d$ has already been inserted. %We note that in general one can use the Kraus representation to describe all completely-positive maps, namely $\Lambda(\rho)=\sum_{i}K_i \rho K_i^\dagger$, with the Kraus operators $K_i$. 
\red{Note that, in general, the success probability depends on the initial state. However, since we consider the maximally mixed state as the input, the postmeasurement states correspond to eigenstates of the von-Neumann measurement operators. Therefore, the success probability depends on the measurement set. Similar to the concept in hidden CHSH nonlocality, the CHSH macrorealistic channel, which can be activated to a CHSH nonmacrorealistic one by SPPOs, is denoted as hidden CHSH nonmacrorealistic channel.}

Physically, $\Lambda$ can be experimentally implemented by weak measurements~\cite{Kim2011}, or a filtering process~\cite{HorodeckiRMP,Horodecki1998,Kwiat2001,Gisin1996,Popescu1995,Ku2021-2}.
% \red{HY: I am still wondering whether the following statement can happy everyone. As Clive said, if we talk about the clumsy measurement, it is super hard too say something beautifully.} 
\red{In the next section, we will explicitly show how filtering operations can activate a hidden CHSH nonmacrorealistic channel.}
We also note that weak measurements have been used to experimentally demonstrate standard nonmacrorealism in different physical systems, including photonic systems~\cite{Goggin2011}, superconducting circuits~\cite{Knee2012}, and atom transport systems~\cite{Robens15}.

Recently, it has been shown that a channel which is not a CHSH nonlocality-breaking channel can be witnessed by violating the temporal CHSH inequality~\cite{Ku2021}. In other words, if the temporal CHSH inequality can be violated, the channel is not CHSH nonlocality-breaking. Here, we further show how temporal quantum correlations can be used to distinguish the strongly CHSH nonlocality-breaking channels from their counterparts~\cite{Pal2015}. %     hidden CHSH nonmacrorealistic channels can be used to verify the non-strongly CHSH nonlocality-breaking channels

\red{We consider the Choi representation of the channel with pre and post filters, namely $\rho_{\text{CJ}}:=(\openone\otimes\Lambda_{\text{post}}\circ \mathcal{E}\circ\Lambda_{\text{pre}})\ket{\Phi_+}\bra{\Phi_+}$, the output state under a general physical map (CP map is up to renormalization) \red{$\Lambda(X)=\text{Tr}_{\text{in}}\left[\openone\otimes X^{\intercal}\rho_{\text{CJ}}\right]$}~\cite{Chiribella2009}, and the property of the maximally entangled state $(\openone\otimes X)\ket{\Phi_+}=(X^{\intercal}\otimes \openone)\ket{\Phi_+}$. Here, $X$ represents any operator and $\intercal$ denotes the transpose. To compare with the strongly CHSH nonlocality-breaking channel, we further assume that the normalization constants are uniform $N(a|x)=N~\forall~a,x$ [cf. Eq.~\eqref{eq:state_after_local_filter}]. 
In other words, the success probability is independent of the choice of the measurement set $\{M_{a|x}\}$ (see also Ref.~\cite{Uola2020} which discusses the same issue in the LGI).
Inserting the above into Eq.~\eqref{Eq_pabxy_instrument}, we can see that testing hidden CHSH nonmacrorealistic channels is the same as certifying whether the Choi state of the channel $\mathcal{E}$ shows hidden CHSH nonlocality (cf., Section~\ref{sec:PRELIMINARY NOTIONS}~\cite{Pal2015}).
Equivalently, if there is hidden CHSH nonmacrorealism, the corresponding channel is not strongly CHSH nonlocality-breaking. Thus, the temporal Bell inequality can be used to certify the quantumness of a quantum channel in a quantum network.}

\section{Experimental demonstration}\label{Sec_Experimental_demonstration}
As a proof-of-principle demonstration, we consider the hidden CHSH nonmacrorealism in an amplitude-damping channel with photonic qubits. The amplitude damping is commonly used to describe the energy dissipation or spontaneous emission. An amplitude-damping channel $\mathcal{E}(\rho)$ can be decomposed into
\begin{equation}\label{Eq_amplitude}
\mathcal{E}(\rho)=E_1 \rho E_1^{\dagger}+E_2\rho E_2^{\dagger},
\end{equation}
with
%\begin{subequations}
%\begin{center}
\begin{eqnarray}
\label{Eamp}
E_1&=&\sqrt{1-v}\ket{1}\bra{1}+\ket{0}\bra{0},  \\
E_2&=&\sqrt{v}\ket{0}\bra{1}, 
\end{eqnarray}
%\end{center}
%\end{subequations} 
corresponding to the decay of the population in $\ket{1}$ and the quantum jump from $\ket{1}$ to $\ket{0}$, respectively, and $v\in[0,1]$ being the visibility. At high visibility, the qubit in the amplitude-damping channel \red{tends} to be in the state $\ket{0}$ and \red{exhibits \red{CHSH} macrorealism ($B_{\rm T-CHSH} \leq 2$)} for $v \geq v_{\rm th} = 0.5$. However, as demonstrated by our experiment, the \red{CHSH} nonmacrorealism \red{($B_{\rm T-CHSH} > 2$)} can be observed \red{for} $v > v_{\rm th}$ if the following SPPO (filtering processes) are exploited,
\begin{eqnarray} 
\label{Eq_SPOOS_Kraus}
\red{K_{\text{pre}}}&=&\ket{0}\bra{0}+\sqrt{1-D}\ket{1}\bra{1},  \\
\red{K_{\text{post}}}&=&\sqrt{1-D}\ket{0}\bra{0}+\ket{1}\bra{1},
\end{eqnarray}
where $D\in[0,1]$ is the power loss of state $\ket{0}$.
%\blue{HY: Change $K_1$ and $K_2$ to $K_{\text{pre}}$ and $K_{\text{post}}$. And describe more from my previous note.} \green{HC: We include the descriptions in the report.}

\subsection{Experimental Setup}
\begin{figure}[tbp]
\includegraphics[width=1\columnwidth]{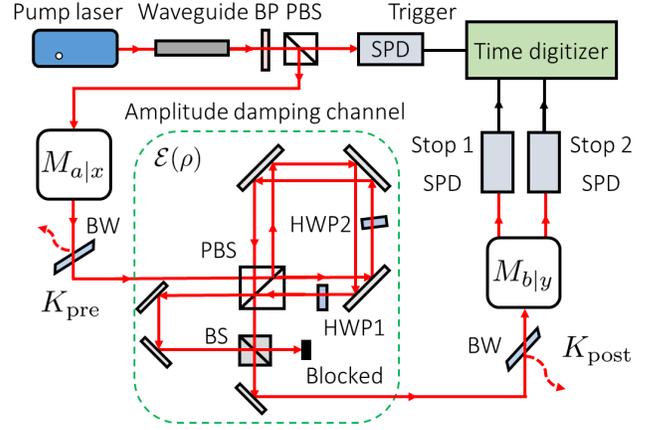}
\caption{Experimental setup. BP: Band-pass filter; SPD: single-photon detection modules. BW: Brewster window plate. The time digitizer is triggered by the horizontally polarized photons. The coincidence counts are recorded once the measurements $M_{b|y}$ lead to clicks at Stop 1 or Stop 2.} 
\label{fig_setup}\centering 
\end{figure}

The schematic of our experimental setup is illustrated in Fig.~\ref{fig_setup}. The type-II quasi-phase-matched spontaneous parametric down-conversion (SPDC) is realized in a periodically-poled Lithium Niobate waveguide pumped by a 780-nm CW laser (ECDL system). The co-propagating 1560-nm photon pairs are spatially separated with a polarizing beam splitter (PBS). This allows us to herald a single photon in the signal channel by detecting an idler photon. To test the temporal CHSH inequality, we send a maximally mixed state into the amplitude-damping channel at $t_0$ by preparing the qubit in an ensemble of the eigenstates of $\hat{\sigma}_x$ and $ \hat{\sigma}_y$ with equal probability \cite{Bartkiewicz2016-2}. In addition, we use the von-Neumann measurement sets $\{M_{a|x}\}=\{\hat{\sigma}_x$,$ \hat{\sigma}_y\}$ at time $t_0$ and $\{M_{b|y}\}=\{(\hat{\sigma}_x+\hat{\sigma}_y)/\sqrt{2}, (\hat{\sigma}_x-\hat{\sigma}_y)/\sqrt{2} \}$ at time $t_1$.

The amplitude-damping channel is implemented in a Sagnac-like interferometer~\red{\cite{Almeida2007,Kim2011} } with polarization encoding, $|0\rangle = |H\rangle$ and $|1\rangle = |V\rangle$. The $|H\rangle$ and $|V\rangle$ components of the qubit are first separated by a PBS. \red{ The $|V\rangle$ component is then passed though a half-wave plate (HWP1), with its fast axis rotated at $0<\theta<\pi/2$ ($\sin{2\theta}=\sqrt{v}$), and converted probabilistically to the $|H\rangle$ component after the PBS. To compensate the additional phase induced by HWP1, another half-wave plate (HWP2) tilted at an angle is inserted in the interferometer arm of the $|H\rangle$ component.} After \red{leaving the interferometer,} the $|H\rangle$ component is temporally delayed by \red{ traveling a longer optical path than that by the $|V\rangle$ component} \red{before the two components are recombined incoherently} on a non-polarizing beam-splitter (BS). By doing so, an initial state like 
\begin{equation}
\rho=(\alpha |H \rangle +\beta |V\rangle)(\alpha^* \langle H| +\beta^* \langle V|)\label{ori_rho}    
\end{equation}
is altered to
\begin{equation}
\begin{aligned}
\mathcal{E}(\rho)&=(\alpha |H\rangle +\beta \cos{2\theta} |V\rangle) (\alpha^*\langle H| 
+\beta^* \cos{2\theta} \langle V|) \\&+\beta^2 \sin^2{2\theta} |H \rangle \langle H|,\label{AD_rho}
\end{aligned}
\end{equation}
in the amplitude-damping channel. Essential to our demonstration is the \red{ $K_{\text{pre}}$ and $K_{\text{post}}$} of the SPPOs. These operators are implemented by the glass windows placed at the Brewster angles. By adjusting the principle axes of the windows, we can filter out the $|V\rangle$ or $|H\rangle$ component, which corresponds to the actions $\Lambda_{\rm{pre}} $ and $\Lambda_{\rm{post}}$ in Fig.~\ref{fig_setup}. In our experiment, $D$ is approximately 0.45 for both \red{ $K_{\text{pre}}$ and $K_{\text{post}}$}.

\subsection{Experimental Results}
\begin{figure}[tbp]
\includegraphics[width=1\columnwidth]{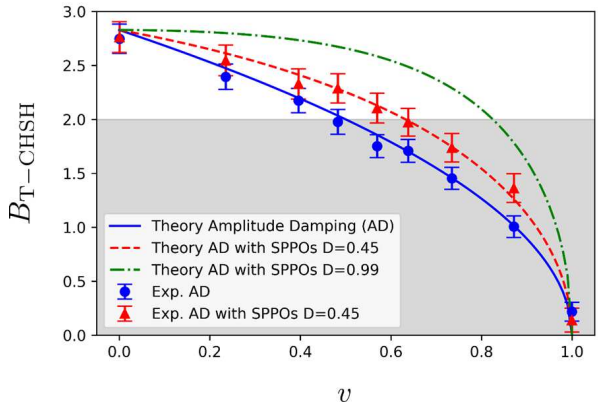}
\caption{Observation of the hidden CHSH nonmacrorealism in an amplitude-damping channel with visibility $v$. The grey shaded area at the bottom represent the macrorealisic regime.
Without applying the SPPOs (blue dots with solid curves being the theoretical prediction), the temporal CHSH inequality is violated $B_{\rm{T-CHSH}} > 2$ up to $v = 0.48$. With the SPPOs (red triangles with dashed curves), the violations are observed up to $v = 0.64$. The green dot-dashed curve represents the theoretical prediction with loss $D \approx 1$ (where $D=0.99$). \red{The error bars take into account the finite accuracy of the waveplate's angle ($\pm 1\degree$), imperfect Brewster windows ($D\pm 2\%$), misaligned incident polarization ($\pm 1 \degree$), and nonideal visibility ($96-98\%$) of the Sagnac interferometer.}}
\label{fig_result}\centering 
\end{figure}

\red{The hidden \red{CHSH} nonmacrorealism is experimentally probed by the violation of the temporal CHSH inequality Eq.~(\ref{Eq_CHSH})}. Figure~\ref{fig_result} shows the measured $B_{\rm T-CHSH}$ versus the visibility $v=\sin^2{2\theta}$. Without the SPPOs (circular dots), we observe the \red{CHSH} nonmacrorealism up to $v = 0.48$, which is in good agreement with the theory (solid curve). The slight deviation between our observation and theory likely results from the \red{finite accuracy of the waveplate's angle, non-ideal Brewster windows, misaligned incident polarization, and imperfect visibility of the Sagnac interferometer}. After applying the SPPOs (triangular dots), the violation of the temporal CHSH inequality can be observed up to $v_{\rm max}=0.64 > v_{\rm th}$, thus demonstrating the hidden \red{CHSH} nonmacrorealism in the region of $0.5 < v \leq 0.64$. We note that the range of the observable hidden \red{CHSH} nonmacrorealism is only limited by the power loss $D$ chosen for the SPPOs in Eq.~\eqref{Eq_SPOOS_Kraus}. For example, the hidden \red{CHSH} nonmacrorealism can be observed up to $v_{\rm max}= 0.83$ if $D \approx 1$, which can be achieved by using more glass windows at the Brewster angles. \red{We also note that when $v>0.5$, the channel is CHSH nonlocality-breaking~\cite{Zhang2020}.} Therefore, our experimental observation implies that the amplitude-damping channel is not a strongly CHSH-nonlocality-breaking channel~\cite{Pal2015} for $0.5 < v \leq 0.64$. 

\section{Discussion}\label{sec:conclusion}
In this work, we have proposed and demonstrated hidden nonmacrorealism. More specifically, we have shown how macrorealistic channels can be ``activated" to nonmacrorealistic channels by using the stochastic pre- and post-operations (SPPOs) both theoretically and experimentally. In other words, our proposal provides a heuristic way to partially mitigate certain types of temporal quantum correlations against decoherence. In addition, we have shown how strongly CHSH nonlocality-breaking channels are related to the hidden CHSH nonmacrorealistic channels. In general, when the system is hidden CHSH nonmacrorealistic, it is not a strongly CHSH nonlocality-breaking channel.

This work also suggests some open problems. How can one determine the set of hidden nonmacrorealistic channels? Is it enough to test whether the Choi state of the channel shows the hidden nonmacrorealistic channels? %In our experiment, we have demonstrated the hidden nonmacrorealism in an amplitude-damping channel for $0.5 < v \leq 0.64$. %As the amplitude-damping channel is not a strongly CHSH nonlocality breaking channel, can we further increase the theoretical bound from $v=0.83$ to $v=1$? 
As we have shown in our experiment, certain consequences of decoherence can be suppressed by applying SPPOs. In fact, similar ideas can be generalized and applied to different problems, e.g., reviving quantum interference~\cite{Chih2019} and quantum teleportation power~\cite{Li2020}.

\section*{Acknowledgments}
The authors acknowledge fruitful discussions with S.-L. Chen, C.-Y. Cheng, J.-Y. Li, and G. N. M. Tabia. H.-Y.Ku acknowledges the support of the Ministry of Science and Technology, Taiwan (Grant No. MOST 110-2811-M-006-546). N.Lambert acknowledges partial support from JST PRESTO through Grant No. JPMJPR18GC.
This work is supported partially by the National Center for Theoretical Sciences and Ministry of Science and Technology, Taiwan, Grants No. MOST 110-2123-M-006-001 and MOST 109-2627-M-006-004, and the Army Research Office (under Grant No. W911NF-19-1-0081).
H.-C. Weng, Y.-A. Shih, and C.-S. Chuu acknowledge the support of the Ministry of Science and Technology, Taiwan (107-2112-M-007-004-MY3). F.N. is supported in part by: Nippon
Telegraph and Telephone Corporation (NTT) Research, the Japan
Science and Technology Agency (JST) [via the Quantum Leap Flagship
Program (Q-LEAP) program, the Moonshot R\&D Grant Number
JPMJMS2061, and the Centers of Research Excellence in Science and
Technology (CREST) Grant No. JPMJCR1676], the Japan Society for
the Promotion of Science (JSPS) [via the Grants-in-Aid for
Scientific Research (KAKENHI) Grant No. JP20H00134 and the
JSPS-RFBR Grant No. JPJSBP120194828], the Army Research Office
(ARO) (Grant No. W911NF-18-1-0358), the Asian Office of Aerospace
Research and Development (AOARD) (via Grant No. FA2386-20-1-4069),
and the Foundational Questions Institute Fund (FQXi) via Grant No.
FQXi-IAF19-06.

% Start appendix
\appendix

%\bibliography{Paper}

%}


\end{document}